\documentclass[aps,prb,twocolumn,noeprint,citeautoscript,superscriptaddress]{revtex4-1}

\usepackage[log-declarations=false]{xparse}
\usepackage{physics}
\usepackage[pdftex]{graphics}
\usepackage{amssymb,amsmath}
\usepackage{subscript}
\usepackage{bm}

\usepackage{color}
\usepackage[usenames,dvipsnames]{xcolor}
\definecolor{myblue}{rgb}{0,0,1}
\usepackage[breaklinks=true,colorlinks=true,linkcolor=myblue,urlcolor=myblue,citecol  or=myblue]{hyperref}

\usepackage{txfonts}

\begin{document}

\title{\emph{Ab Initio} Linear and Pump-Probe Spectroscopy of Excitons in Molecular Crystals}

\author{Alan M.~Lewis}
\affiliation{Department of Chemistry and James Franck Institute,
University of Chicago, Chicago, Illinois 60637, United States}

\author{Timothy C.~Berkelbach}
\email{tim.berkelbach@gmail.com}
\affiliation{Department of Chemistry, Columbia University, New York, New York 10027, United States}
\affiliation{Center for Computational Quantum Physics, Flatiron Institute, New York, New York 10010, United States}

\begin{abstract}
Linear and non-linear spectroscopies are powerful tools used to investigate the
energetics and dynamics of electronic excited states of both molecules and
crystals. While highly accurate \emph{ab initio} calculations of molecular
spectra can be performed relatively routinely, extending these calculations to
periodic systems is challenging. Here, we present calculations of the linear
absorption spectrum and pump-probe two-photon photoemission spectra of the
naphthalene crystal using equation-of-motion coupled-cluster theory with single
and double excitations (EOM-CCSD).  Molecular acene crystals are of interest due
to the low-energy multi-exciton singlet states they exhibit, which have been
studied extensively as intermediates involved in singlet fission.  Our linear
absorption spectrum is in good agreement with experiment, predicting a first
exciton absorption peak at 4.4~eV, and our two-photon photoemission spectra
capture the behavior of multi-exciton states, whose double-excitation character
cannot be captured by current methods.  The simulated pump-probe spectra provide
support for existing interpretations of two-photon photoemission in
closely-related acene crystals such as tetracene and pentacene.
\end{abstract}

\maketitle

Multiphoton spectroscopies such as two-photon photoemission spectroscopy are
increasingly being applied to molecules, clusters, and solids as a complement to
linear
spectroscopies.\cite{Neumark2001,Lu2008,Chan2011,Chan2012,Faure2012,Ramasesha2016}
These spectroscopies allow the direct investigation of the character and
dynamics of excited states, in contrast to the ground state properties probed by
linear spectroscopies.  From a computational perspective, the \textit{ab initio}
simulation of both linear and nonlinear spectroscopies is an ongoing challenge,
especially for solid-state systems, which limits the interplay between theory
and experiment.

The current state of the art for the simulation of linear spectra of
semiconductors and insulators with excitonic effects is based on Green's
functions, in particular the Bethe-Salpeter equation (BSE) based on the GW
approximation to the
self-energy~\cite{Hedin1965,sham_many-particle_1966,hanke_many-particle_1980,strinati_dynamical_1980,strinati_dynamical_1982,strinati_effects_1984,
Hybertsen1985,Hybertsen1986,albrecht_excitonic_1998,Rohlfing2000,Ridolfi2018}.
However, the extension of these methods to \textit{ab initio} nonlinear
spectroscopies is not
straightforward~\cite{Chang2001,Perfetto2015,Attaccalite2018} and the treatment
of double excitations is difficult (and impossible within the adiabatic
approximation of time-dependent density functional theory or the common static
screening approximation to the BSE)~\cite{Giesbertz2008,Romaniello2009}.

By contrast, wavefunction-based quantum chemistry techniques are regularly
employed to calculate the properties and spectra of molecules with high
accuracy, and simulating nonlinear spectroscopies with the inclusion of double
excitations is achievable.  For example, in recent years, quantum chemistry
techniques to simulate two-photon photoemission spectra have been
developed\cite{Oana2007,Gozem2015} and used to study a range of molecular
phenomena, such as the S$_2$/S$_1$ conical intersection in
benzene,\cite{Thompson2011} the electronic states of the unpaired electron in
sodium clusters,\cite{Gunina2016} and the ring opening of
1,3-cyclohexadiene.\cite{Tudorovskaya2018}

Bringing the predictive capabilities of coupled-cluster theory to bear on
solid-state problems with explicit periodic boundary conditions is difficult
because of the comparatively high cost and large system sizes required to make
predictions near the thermodynamic limit.  While early calculations demonstrated
the promise of this approach~\cite{Sun1996,Hirata2004}, only recently have
periodic perturbation theory\cite{Ayala2001,Pisani2005,Marsman2009} and
coupled-cluster calculations been performed for
ground-state~\cite{Booth2013,Gruneis2015,McClain2017,Gruber2018} and
excited-state~\cite{McClain2016,McClain2017,Lewis2019} properties of
three-dimensional systems.

\begin{figure*}
\centering
\resizebox{1.9\columnwidth}{!} {\includegraphics{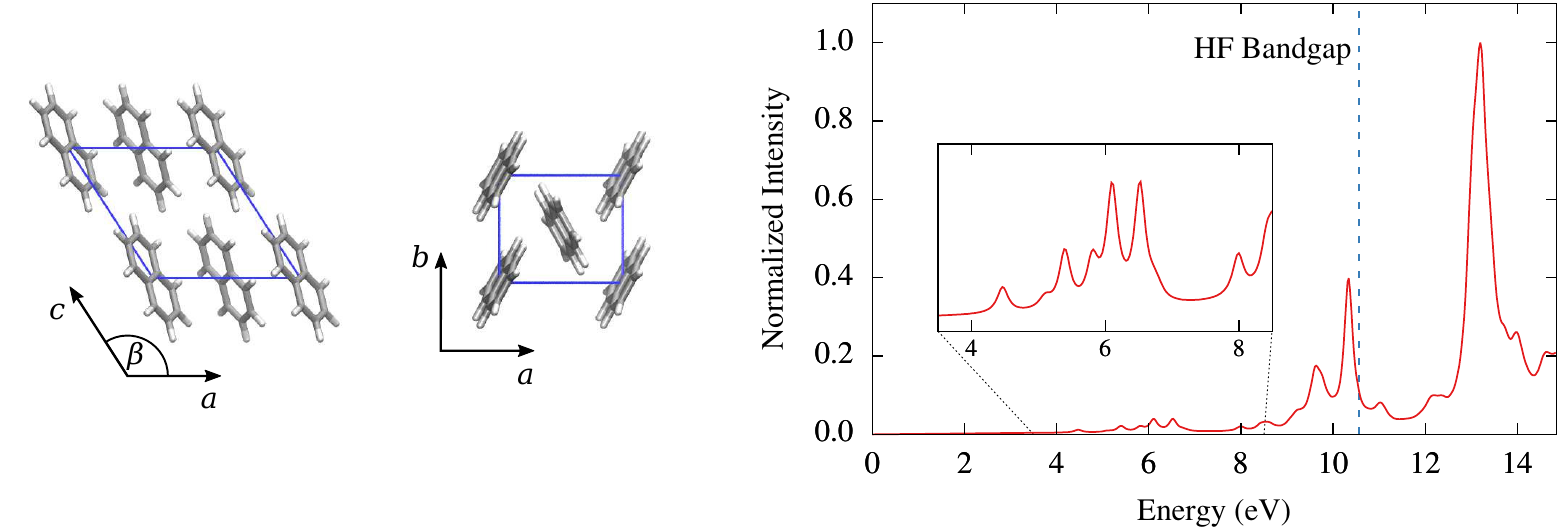}}
\caption{Left: The unit cell of naphthalene; $a = 8.235$ \AA, $b = 6.003$ \AA,
$c = 8.658$ \AA, $\beta = 123^\circ$. Crystallographic data was taken from
Abrahams \textit{et al.}\cite{Abrahams1949} Right: The polarization-averaged
linear absorption spectrum of the naphthalene crystal calculated using
EE-EOM-CCSD and Hartree-Fock. }
\label{unit_cell_LinAbs}
\end{figure*}

Here we build on these developments, presenting both the linear absorption
spectrum and two-photon photoemission spectra of a molecular crystal calculated
using coupled-cluster theory with single and double excitations (CCSD). In
particular, we calculate spectra of the naphthalene crystal, whose unit cell is
shown in Figure \ref{unit_cell_LinAbs}.\cite{Abrahams1949} Like other acene
crystals, naphthalene exhibits a low-lying multi-exciton state -- an overall
singlet state that is qualitatively composed of two triplet excitons on
neighboring molecules.\cite{Smith2010,Stern2015,Miyata2019} We shall refer to
this state as S$_{\rm ME}$ throughout. Multi-exciton states are of particular
interest as precursors to singlet fission, since in principle they allow a
single photon to generate two free
charges.\cite{Smith2010,Chan2011,Chan2012,Smith2013,Congreve2013,Zhu2018,Kim2018}
The energetics\cite{TuanTrinh2017,Tempelaar2017,Folie2018} and
dynamics\cite{Wan2015,Monahan2017,Tempelaar2018} of these multi-exciton states
are the subject of considerable ongoing research.

The remainder of the paper is organized as follows. We first calculate the
neutral excited states of the naphthalene crystal using periodic
equation-of-motion CCSD for electronic excitations (EE-EOM-CCSD) and simulate
the linear absorption spectrum, which is compared to experiment.  Then we
calculate the ionization spectra of the ground state and various excited states,
the latter of which approximates the two-photon photoemission spectrum (2PPE).
We compare to experimental 2PPE spectra on related acene crystals and comment on
the signatures of multi-exciton character.

\section*{Results and discussion}

\subsection*{Neutral excitations and linear absorption}

In order to investigate excitonic and multi-excitonic states of the naphthalene
crystal, we use EE-EOM-CCSD, which is well suited to calculating these excited
states in periodic systems since it produces a size-extensive total energy and
size-intensive excitation energies.\cite{Bartlett2007} Our calculations are
performed on a single unit cell at the gamma point with periodic boundary
conditions using pseudopotentials and a double-zeta basis set.  Further details
and a discussion of finite-size effects can be found in Methods.  In
Table~\ref{EEs}, we compare the excitation energies of the naphthalene crystal
to those of the naphthalene monomer and a dimer in the crystal phase geometry.
Despite the differences between the molecular and periodic calculations, we
observe very similar excitation energies in the monomer, dimer, and the crystal.
This behavior is consistent with a picture of tightly-bound excitons in
molecular crystals.

The calculated energy of the S$_{\rm ME}$ state is roughly 8~eV in both the
dimer and the crystal, which is significantly more than twice the T$_1$ energy.
This behavior is a consequence of the well-known tendency of EOM-CCSD to
overestimate the energy of states dominated by double
excitations~\cite{Stanton1993,Kowalski2004,Kim2018} due to a lack of orbital
optimization.  Nonetheless, even the qualitative description of multiexciton
states is still an outstanding challenge for alternative techniques such as the
GW-BSE approach~\cite{Onida2002,Romaniello2009,Coto2015}, and we will later show
how this state can still be used for a qualitative -- and even quantitative --
understanding of multiexciton physics.  Within molecular quantum chemistry,
encouraging results have been obtained for the multiexciton state of acenes
using multireference active space methods~\cite{Abraham2017,Chien2017,Yong2017}.
We consider this an important area for future work on periodic systems.

\begin{table}
\begin{tabular}{ c | c | c | c }
 & S$_1$ Energy & T$_1$ Energy & S$_{\rm ME}$ Energy \\
\hline
Monomer & 4.5 & 3.1 & - \\
Dimer & 4.5 & 3.1 & 7.6 \\
Crystal & 4.4 & 3.3 & 8.3 
\end{tabular}
\caption{Selected excitation energies of the naphthalene monomer, dimer, and
crystal calculated using EE-EOM-CCSD. S$_{\rm ME}$ refers to the lowest-energy
singlet multi-exciton state, characterized by a low quasiparticle weight. All
energies are in eV.}
\label{EEs}
\end{table}

In Fig.~\ref{unit_cell_LinAbs}, we present the polarization-averaged linear
absorption spectrum of the naphthalene crystal
\begin{equation}
S(\omega) = \sum_{\mu>0} \sum_{\lambda=(x,y,z)} \langle \tilde{\Psi}_0 | \hat{P}_\lambda |\Psi_\mu\rangle 
    \langle \tilde{\Psi}_\mu | \hat{P}_\lambda |\Psi_0\rangle \delta(\omega - (E_\mu-E_0))
\label{LinAbsEq}
\end{equation}
where $\Psi_0$ ($\tilde{\Psi}_0$) is the right-hand (left-hand) CCSD ground
state with energy $E_0$ and $\Psi_\mu$ ($\tilde{\Psi}_\mu$) is a right-hand
(left-hand) EE-EOM-CCSD excited state with energy $E_\mu$.  In the periodic
setting, the transition strength is determined by matrix elements of the
components of the momentum operator $\hat{P}_\lambda = \sum_{pq}
P^{(\lambda)}_{pq}\hat{a}_p^\dagger \hat{a}_q$, where
$\hat{a}^\dagger_p$ ($\hat{a}_q$)
creates (annihilates) an electron in molecular orbital $p$ ($q$).  More details
are provided in Methods.

At low energies, the spectrum is dominated by narrow peaks below the band gap,
signaling the presence of excitons.  The first peak appears at the S$_1$ energy
of 4.4~eV, as reported in Table~\ref{EEs}.  This value is in good agreement with
a previous BSE calculation (3.9~eV) and with experimental values
(3.9-4.0~eV).\cite{Silinsh1980,Pope1999,Hummer2005} As indicated in
Fig.~\ref{unit_cell_LinAbs}, the Hartree-Fock (HF) bandgap is over 10~eV, which
demonstrates that electron correlation makes a large contribution to the optical
excitation energy.  At higher energies, the spectrum becomes more broad and more
intense due to a combination of the greater density of states and larger
transition matrix elements.  Due to the one-body nature of the momentum
operator, the linear absorption spectrum primarily reports on excited states
with predominant single excitation character, i.e.~excitons and interband
transitions.  Having established the quality of singly-excited states predicted
by periodic EOM-CCSD, we now turn to the simulation of pump-probe spectroscopy,
which can report on states with predominant double excitation character.

\subsection*{Two-Photon Photoemission}

\begin{figure*}
\centering
\resizebox{1.9\columnwidth}{!} {\includegraphics{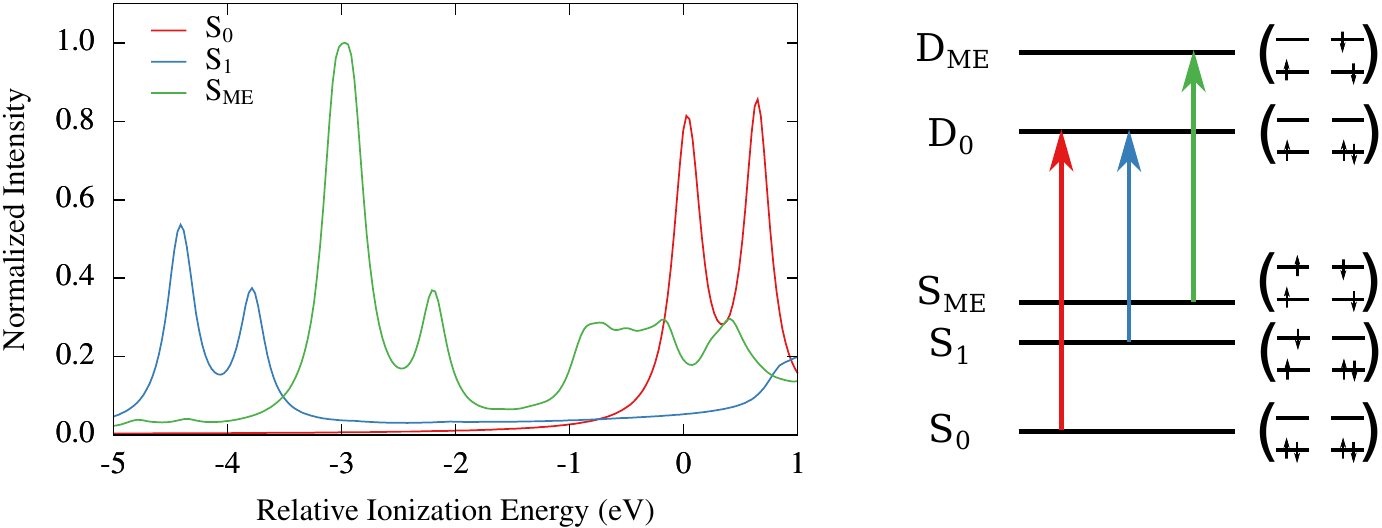}}
\caption{Left: The low-energy part of the photoemission spectrum from the S$_0$, S$_1$
and S$_{\rm ME}$ states of naphthalene, calculated using a combination of IP-
and EE-EOM-CCSD. Right: A simplified molecular orbital diagram of two naphthalene molecules in
the crystal, which qualitatively explains the calculated photoemission spectrum
from the S$_0$, S$_1$ and S$_{\rm ME}$ states.}
\label{fig:2PPE}
\end{figure*}

In order to simulate the two-photon photoemission spectrum, we calculate the
total ionization spectrum of each neutral excited state $\Psi_\mu$,
\begin{equation}
\begin{split}
\label{eq:2PPE}
A^{\mathrm{(2PPE)}}_\mu(\omega) &= \sum_{\nu,p} \mel{\tilde{\Psi}^N_\mu}{\hat{a}_p^{\dagger}}{\Psi^{(N-1)}_\nu} 
    \mel{\tilde{\Psi}^{(N-1)}_\nu}{\hat{a}_p}{\Psi^N_\mu} \\
    &\hspace{1em} \times \delta(\omega - (E^{(N-1)}_\nu - E^N_\mu)).
\end{split}
\end{equation}
The notation is as above, except we emphasize that the final states are ionized
states with $N-1$ electrons, calculated using EOM-CCSD for ionization potentials
(IP-EOM-CCSD).  This signal can be thought of as the trace of the imaginary part
of an \textit{excited-state} one-particle Green's function.  For $\mu=0$, this
gives the usual ground-state one-particle spectral function or many-body density
of occupied states, which is a common approximation to the photoemission
signal~\cite{Thouless1972,Cederbaum1975,Abrikosov2012}.  Information about the
intensity of the signal can be included with an appropriate matrix
element~\cite{Oana2007,Gozem2015}. However this depends on details of the
experiment being modeled and, in the periodic setting, details of the surface
termination.  Physically, the above expression models the scenario that a pump
pulse has prepared the excited state $\Psi_\mu$ or, alternatively, that a pump
pulse has prepared a nonstationary distribution that nonadiabatically evolves
into the excited state $\Psi_\mu$.  Although our calculation is not
time-resolved and neglects electronic and nuclear dynamics, the interpretation
of time-resolved spectra in terms of time-independent state-specific spectra is
a common approach in the analysis of experimental transient
data~\cite{VanStokkum2004}.

The ionization spectrum of the ground state S$_0$, first singlet excited state
S$_1$, and first singlet multi-exciton state S$_\mathrm{ME}$ of the naphthalene
crystal are shown in Fig.~\ref{fig:2PPE}. Ionization energies are plotted
relative to the first ionization energy of the ground state such that a negative
value indicates that less energy is required for ionization, which would leave
more kinetic energy in a photoemission experiment. 
All of the spectra exhibit peaks in pairs, arising from the hybridization of the
orbitals of the two inequivalent molecules in the unit cell, akin to the
well-known Davydov splitting~\cite{Davydov1948}. The first ionization energy of
S$_1$ is significantly lower than that of S$_0$; as can be understood from
Eq.~\ref{eq:2PPE}, the difference between these peaks is precisely the S$_1$
excitation energy of 4.4~eV.  This suggests a general trend that higher-lying
neutral excited states will have a first ionization peak at increasingly
negative relative energies.  However, the first significant peak in the
ionization spectrum of S$_{\rm ME}$ is higher in energy (less negative) than the
corresponding peak in the S$_1$ spectrum, despite the fact that the excitation
energy of S$_{\rm ME}$ is roughly twice that of S$_1$. 

This unexpected ordering can be understood as a final state effect, shown
schematically in Fig.~\ref{fig:2PPE} for two neighboring naphthalene molecules.
The lowest energy ionization of either the S$_0$ or S$_1$ state produces the
same final state, the ground state of the ion labeled D$_0$. As a result, the
difference in the first ionization energy of these states is simply the
excitation energy of S$_1$. By contrast, ionizing S$_{\rm ME}$ produces an
\textit{excited} state of the ion, which we label D$_{\rm ME}$.  As a result,
the ionization energy is larger than one might naively predict based on the
excitation energy of the S$_{\rm ME}$ state alone.

In the simple two-molecule picture, ionization of S$_{\rm ME}$ produces a state
that can be thought of as one ionized molecule and one molecule in its first
triplet state, such that $E({\rm D}_{\rm ME}) - E({\rm D}_0) \approx E({\rm
T}_1)$.  Therefore, the first ionization peak of S$_{\rm ME}$ should be shifted
from that of S$_0$ by about $E({\rm T}_1) - E({\rm S}_{\rm ME}) \approx -E({\rm
T}_1)$, which is consistent with our observed shift of about $-3$~eV.

In the language of IP-EOM-CCSD, the D$_{\rm ME}$ state is also a double
excitation, like S$_{\rm ME}$, corresponding to a two-hole+one-particle
excitation, for which the absolute value of the energy of the D$_{\rm ME}$ state
is surely overestimated.  However, the energy difference between two states with
predominant double excitation character, D$_{\rm ME}$ and S$_{\rm ME}$, is
likely to benefit from a cancellation of errors, since neither includes orbital
optimization.  Therefore, we expect that the 2PPE spectrum of all states shown
in Fig.~\ref{fig:2PPE} is quite accurate.

This counter-intuitive ordering of the first ionization energies of S$_1$ and
S$_{\rm ME}$ has been observed experimentally in 2PPE studies of
tetracene~\cite{Chan2012} and pentacene~\cite{Chan2011}. In these cases, the
S$_0$ and S$_{\rm ME}$ state are nearly resonant, yet the ionization energy of
the S$_0$ is approximately 0.7 -- 1~eV lower than the ionization energy of
S$_{\rm ME}$. These values are approximately the energy of a triplet exciton in
tetracene and pentacene, in agreement with our simulated result on naphthalene.
Taken together, these observations suggest that this ordering of the ionization
energies is characteristic of multi-exciton states and that two-photon
photoemission spectroscopy is well-suited to identifying these
states.\cite{Chan2011,Chan2012}   

\section*{Conclusions}

We have performed calculations of the linear absorption and pump-probe
two-photon photoemission spectrum of the napthalene crystal using
equation-of-motion coupled-cluster theory with explicit periodic boundary
conditions.  Our results demonstrate an accurate description of low-lying
excitons in molecular crystals.  Importantly, the ability to describe
wavefunctions with double excitation character provides access to multiexciton
states that are relevant for technologically important processes such as singlet
fission and that can be probed via pump-probe spectroscopies.  In addition to
providing an \textit{ab initio} description of recent experimental results on
related molecular acene crystals, our work establishes molecular quantum
chemistry techniques, such as coupled-cluster theory, as promising methods for a
description of the excited-state electronic structure of solids.

\section*{Methods}

All of our calculations were performed using the PySCF software
package~\cite{pyscf}, with the exception of the density functional theory
geometry optimization of the naphthalene molecule, which was carried out using
Gaussian09~\cite{gaussian}.  Our CCSD calculations use a Hartree-Fock reference
state calculated using the cc-pvdz basis set for the naphthalene monomer and
dimer and the GTH-DZVP basis set~\cite{Vandevondele2005} for the crystal.  In
post-HF steps, the core occupied orbitals were frozen in the molecular
calculations and GTH-PADE pseudopotentials~\cite{Goedecker1996} were used in the
crystal calculations.  Furthermore, we correlated 20 virtual orbitals per
molecule, which we found to be sufficient to converge the excitation energies to
0.1~eV.

Periodic integrals were evaluated via an auxiliary plane-wave
basis~\cite{McClain2017} with a kinetic energy cutoff of 70 Hartree.  In this
basis, the Coulomb kernel $v(\bm{G})$ is divergent when $\bm{G}=0$ and in all
two-electron integrals we replace this divergence by the Madelung constant
according to the probe-charge Ewald technique~\cite{Sundararaman2013}, except in
the evaluation of the Hartree potential where the divergence is exactly canceled
by the electron-nuclear interaction.  All periodic calculations were performed
at the gamma point, which yields a finite-size error. We note that the minimum
band gap occurs at the D point and not the gamma point.  However, the bands of
naphthalene are only weakly dispersive,\cite{Hummer2005b,Fedorov2011} and our
calculations are simplified at the gamma point due to the use of real integrals.

In the periodic setting with nonlocal pseudopotentials $\hat{V}_{\rm nl}$, the
dipole matrix elements are given by~\cite{Baroni1986}
\begin{equation}
\mel{p}{\hat{\bf R}}{q} = \frac{-i \mel{p}{\hat{\bf P}}{q} - \mel{p}{[\hat{V}_{\rm nl},\hat{\bf R}]}{q}}{\varepsilon_p - \varepsilon_q}.
\label{DipMo}
\end{equation}
For simplicity in our calculations, we have neglected the second term in
Eq.~\eqref{DipMo}, which we do not expect to qualitatively modify the linear
absorption intensities. 

\section*{Acknowledgments}

We thank Sivan Refaely-Abramson for helpful discussions concerning GW-BSE
calculations for napthalene.  All calculations were performed using resources
provided by the University of Chicago Research Computing Center and the Flatiron
Institute.  This work was supported by the Air Force Office of Scientific
Research under award number FA9550-18-1-0058.  The Flatiron Institute is a
division of the Simons Foundation.

\end{document}